\documentclass[fleqn,twoside]{article}
\usepackage{espcrc2}
\usepackage{graphicx}
\usepackage{amssymb}

\newcommand{\be}{\begin{eqnarray}}
\newcommand{\ee}{\end{eqnarray}}
\newcommand{\1}{\mathbb{I}}
\newcommand{\tr}{\mathrm{Tr}}
\newcommand{\re}{\mathrm{Re}}
\renewcommand{\det}{\mathrm{Det}}
\newcommand{\el}{\nonumber\\}

\title{Taste breaking in staggered fermions from random matrix theory}

\author{James C. Osborn\address{Physics Department, University of Utah, Salt Lake City, UT 84112, USA}}

\begin{document}

\begin{abstract}
We discuss the construction of a chiral random matrix model for
staggered fermions.
This model includes $O(a^2)$ corrections to the continuum limit of
staggered fermions and is related to the zero momentum limit of the
Lee-Sharpe Lagrangian for staggered fermions.
The naive construction based on a specific expansion in lattice
spacing ($a$) of the Dirac matrix produces the term which gives
the dominant contribution to the observed taste splitting in the pion masses.
A more careful analysis can include extra terms which are also consistent
with the symmetries of staggered fermions.
Lastly I will mention possible uses of the model including studies of
topology and fractional powers of the fermion determinant.
\end{abstract}

\maketitle

%\section{Introduction}

Staggered fermions are a popular way to include quarks on the lattice.
A single staggered fermion determinant in the continuum limit actually
describes four {\em tastes} of quarks, so named to distinguish them
from physical flavors.
At finite lattice spacing the tastes interact and the continuum $SU(4)$
taste symmetry is broken.
The size of the taste breaking can be reduced by using improved actions,
but for current lattice spacings the effects are clearly present in
chiral fits \cite{ChiLog}.

The form of the taste breaking terms for a single flavor (four tastes)
of staggered fermions was worked out as corrections
to the chiral Lagrangian by Lee and Sharpe \cite{LS}.
They included all terms of $O(a^2)$ that are consistent with the
symmetries of staggered fermions.
This was later extended to arbitrary flavor content (most notably 2+1 flavors)
by Bernard and Aubin \cite{BA}.

Here we will outline how to include taste breaking in a
chiral random matrix model.
This model should be equivalent to low energy lattice QCD with
staggered fermions and can be used to study properties of the lattice theory
which are sensitive to the smallest eigenvalues of the Dirac matrix.
A more detailed study of the staggered matrix model and its applications
will appear elsewhere.

\section{Random Matrix Theory}

Chiral random matrix theory has been used extensively to model the low energy
eigenmodes of the Dirac operator.
Its construction is based solely on the symmetries of the Dirac operator.
In a chiral basis the standard Dirac operator for a single quark flavor
takes the form
\be
{\cal D} = 
\left(
 \begin{array}{cc}
  m~ \1_N & i \, W \\
  i \, W^\dagger     & m~ \1_N \\
 \end{array}
\right)
\ee
where $W$ is a complex $N\times N$ matrix,
$\1_N$ is the $N\times N$ identity matrix and $m$ is the quark mass.
Here we will only consider the case of zero topological charge and
three or more colors which corresponds to the Dirac matrix given above.
Due to universality the partition function near the chiral limit is not
sensitive to the exact form for the distribution of the various
matrix elements.
One can therefore make the simplest choice of an independent Gaussian
measure for each matrix element
\be
\mu(W) = \exp \, (-\alpha \, N \, \tr \, W^\dagger W \, ) ~.
\ee
The parameter $\alpha$ is related to the chiral limit of the chiral
condensate ($\Sigma$) and the volume ($V$) by
$\sqrt{\alpha} = V \Sigma / 2 N$.
For a recent review of chiral random matrix models see \cite{VW} and
its references.

In order to extend chiral random matrix theory to include taste breaking
we need to examine the form of the taste breaking interactions.
The original motivation for this work came from examining the classical
expansion of the staggered Dirac operator near the continuum limit given by
Kluberg-Stern et. al. \cite{KS}.
Their expansion starts
\be
  (\gamma_\mu \otimes \1_4) \, D_\mu + ( \1_4 \otimes \1_4 ) \, m
  + a \, (\gamma_5 \otimes \xi_{\nu 5}) \, D_\nu^2 ~.
\label{ksform}
\ee
The notation $(S \otimes T)$ has the standard meaning of the outer
product of a $4\times 4$ spin matrix $S$ with a $4\times 4$ taste matrix
$T$ with $\xi_{\nu 5} = \xi_{\nu} \xi_{5} = \gamma_\nu^* \gamma_5^*$.
The first part is just the standard Dirac operator for four identical
flavors of mass $m$.
The remaining term is suppressed by a factor of the lattice spacing and
breaks the $SU(4)$ taste symmetry.
There is also an $O(ag)$ term which for our purposes is similar to the
included taste breaking term and will be ignored.
The remaining corrections are at least $O(a^2)$.
This form of the expansion is specific to the particular basis chosen
for the continuum Dirac fermions.
In fact Luo \cite{Luo} has shown that with an improved basis the $O(a)$
terms can be removed giving corrections only at $O(a^2)$.
This ambiguity in basis to use for the continuum expansion can be avoided
by instead starting with the same four-fermion operators in the continuum
effective Lagrangian used by Lee and Sharpe \cite{LS}.
Using a Hubbard-Stratonovitch transformation these terms can split into
fermion bilinears which can be used to construct a matrix theory with
the correct symmetries.
We do not present the details of this construction here but will refer
to it later in the paper.

For simplicity here we will continue to construct the matrix model based
on the expansion (\ref{ksform}) that was given.
The taste breaking term in the chiral basis has the form
\be
{\cal T} = ( A_\nu \otimes \sigma_3 ) \otimes \xi_{\nu 5}
\label{tb}
\ee
where $A_\nu$ is an $N \times N$ Hermitian matrix 
and $\sigma_3$ a Pauli matrix.
We again make the simplest choice for the integration measure over
the matrices $A_\nu$ which is that of independent Gaussian matrix elements
given by
\be
\mu(A_\nu) = \exp \, (-\beta \, N \, \tr \, A_\nu^2 \, )
\ee
with some unknown parameter $\beta$.
The partition function is then
\be
Z = \int \mu(W) \, dW \prod_\nu \mu(A_\nu) \, dA_\nu ~ \det( {\cal M} )
\label{stagrmt}
\ee
with
\be
{\cal M} ~=~ {\cal D} \otimes \1_4 ~+~ a \; {\cal T} ~.
\ee
We now will relate this model to the staggered chiral Lagrangian given
by Lee and Sharpe.

\section{Chiral Lagrangian}

\begin{figure}[t]
\begin{center}
\includegraphics[width=0.45\textwidth]{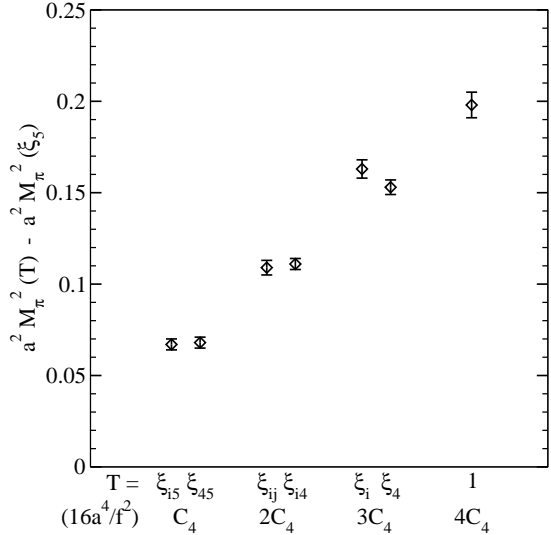}
\vspace{-10mm}
\caption{
Pion mass splittings for different taste content relative to the Goldstone
pion.  The contribution from the $C_4$ term of the staggered chiral
Lagrangian is given at the bottom.
\label{FIGms}
}
\end{center}
\vspace{-8mm}
\end{figure}

The random matrix model (\ref{stagrmt}) can be transformed into a
nonlinear sigma model using standard methods.
In the large $N$ limit the sigma model can be expanded around its
saddle point which gives
\be
{\int}_{U(4)} \, dU \,
 \exp \, (-\,m\,V \, \Sigma \, \re \, \tr \, U \, - \, a^2 \, {\cal V} \, )
\ee
with
\be
{\cal V} ~=~ 
 N \, \frac{\alpha}{\beta} \,
  {\sum}_\nu \, \re \, \tr \, ( \, U \, \xi_{\nu 5} \, U \, \xi_{5\nu} \, ) ~.
\ee
This is equivalent to the zero momentum limit of an effective chiral theory.
Notice that the taste breaking is $O(a^2)$ in the chiral Lagrangian
even though we started with a term of $O(a)$ in the Dirac operator.
The form of ${\cal V}$ is identical to the term with coefficient $C_4$
in the Lee-Sharpe Lagrangian if we set $N \alpha/\beta = V C_4$.
It is interesting to note that this term was previously found to be the
dominant contributor to the mass splittings between pions with different
taste structures.
In Figure \ref{FIGms} we show the difference between the squares of various
pion masses and the square of the Goldstone pion ($\xi_5$) mass in
units of the lattice spacing.
On the $x$-axis are the taste generators used with the corresponding
contribution to the splitting from the $C_4$ term written below.
The data are from simulations of improved ``$a^2$-tad'' staggered fermions at
$\beta_{imp} = 7.3$ and $a m = 0.02$ tabulated in \cite{OT}.
The results agree very well with the linear rise predicted from the $C_4$ term.

We have so far considered only the taste breaking term given in
(\ref{ksform}).
A more general construction alluded to earlier gives the following
additional terms for the matrix model
\be
&( A^{(1)}_\nu \otimes \sigma_{0,3} ) \otimes i \, \xi_{\nu} &\to ~~ C_3 \el
&( A^{(2)} \otimes \sigma_{1} ) \otimes i \, \1_4            &\to ~~ const. \el
&( A^{(3)} \otimes \sigma_{1} ) \otimes i \, \xi_{5}          &\to ~~ C_1 \el
&( A^{(4)}_{\mu\nu} \otimes \sigma_{1} ) \otimes \xi_{\mu\nu} &\to ~~ C_6 ~.
\label{rmmextra}
\ee
The $\sigma_{0,3}$ can be either $\sigma_0 = \1_2$ or $\sigma_3$.
Also the $\sigma_3$ in (\ref{tb}) can be replaced with $\sigma_0$ to
produce another valid term.
The arrow shows which term from the chiral Lagrangian the matrix model
term produces in the large $N$ limit.
The corresponding terms in the chiral Lagrangian are
\be
&& C_1 ~ \tr \; ( \, \xi_{5} \, U \, \xi_{5} \, U^\dagger \, ) \el
&& C_3 ~ {\sum}_\mu ~\re\;\tr\;( \, \xi_{\mu} \, U \, \xi_{\mu} \, U \, ) \el
&& C_6 ~ {\sum}_{\mu<\nu} ~ \tr \;
         ( \, \xi_{\mu\nu} \, U \, \xi_{\nu\mu} \, U^\dagger \, ) ~.
\label{C136}
\ee
The two remaining terms from \cite{LS} are
\be
&& C_2 ~ \re \left[ \, \tr \, (\,U^2\,) \,-\,
  \tr\,(\,\xi_{5}\,U\,\xi_{5}\,U\,)\, \right] \\
&& C_5 ~ \frac12 {\sum}_\mu \, \tr(\,\xi_{\mu}\,U\,\xi_{\mu}\,U^\dagger\,)
                      - \tr\,(\,\xi_{\mu5}\,U\,\xi_{5\mu}\,U^\dagger\,)
. \nonumber
\label{C25}
\ee
They can also be produced from the matrix model using terms similar
to those in (\ref{rmmextra}), however some of the terms would have to
be Hermitian (note the terms in (\ref{rmmextra}) are all anti-Hermitian)
which would make them not suitable for modeling the Dirac matrix.
This point requires further study.

\section{Conclusions and Future Applications}

We have presented a chiral random matrix model for staggered fermions.
In its simplest form (\ref{stagrmt}) it includes just the term that is
known to be dominant in the pion mass splittings.
Additional terms consistent with the staggered fermion symmetries can
also be included, but the construction of the most general form of the
matrix model still requires further study.

This model can also be extended to include additional flavors and
non-zero topological charge.
Both of these extensions will be covered in a future publication.
Since topology is related to the lowest eigenvalues of the Dirac
matrix, this should prove to be a useful model for studying the
topology of staggered fermions.

Current lattice simulations using staggered fermions typically take
the square root or fourth root of the fermion determinant to reduce
the number of tastes from four to produce one or two physical flavors
in the continuum limit.
The effects of the root on the low energy spectrum are not well understood.
One could study its effects in the matrix theory where analytic results
are easier to produce.
If the distribution of Dirac eigenvalues in the matrix model agrees
with lattice calculations, then the effect of the root should be
modeled properly by the matrix theory.
This is currently perhaps the most interesting application of the
staggered random matrix model.

I would like to thank Claude Bernard, Carleton DeTar and Urs Heller for
discussions which helped me understand taste breaking in
staggered fermions better.
This work was funded by by the US NSF.


\begin{thebibliography}{9}
\bibitem{ChiLog} C. Aubin, et. al., Nuc. Phys. B (Proc. Suppl.) 119 (2003) 233.
\bibitem{LS} W. Lee and S.R. Sharpe, Phys. Rev. D 60 (1999) 114503.
\bibitem{BA} C. Bernard, Phys. Rev. D 65 (2002) 054031;
             C. Aubin and C. Bernard, Phys. Rev. D 68 (2003) 034014.
\bibitem{VW} J.J.M. Verbaarschot and T. Wettig, Ann. Rev. Nucl. Part. Sci. 50
             (2000) 343.
\bibitem{KS} H. Kluberg-Stern, et. al., Nuc. Phys. B 220 (1983) 447.
\bibitem{Luo} Y. Luo, Phys. Rev. D 55 (1997) 353;
              Y. Luo, Phys. Rev. D57 (1998) 265.
\bibitem{OT} K. Orginos and D. Toussaint, Phys. Rev. D 59 (1999) 014501.
\end{thebibliography}
\end{document}